# "Are you an AI?" Analyzing client suspicion of AI use in crisis counseling

Shreya Shah[1], Akshay Swaminathan, BA[2], Meghana Simhadri[3], Ivan Lopez, MS[2], Sharang Phadke, BSE[4], Divyanjali Verma, BA, MA, MPhil[5], Abhay John, MBBS, MD, MBA[5], Luke Zhao[6], Fiona Cai, BS, MEng[2], Sharon Zhang, BA[7], Gloria Ye, BS[8,14], Ivy Pham, BA[9], William Wang, MS[10,11], Sebastian Garcia, BA[12], Sarah Wornow, BS[13], Angelina Wang, BS, MA, PhD[14], Nigam H. Shah, MBBS, PhD[8,15,16]

[1] Olentangy Liberty High School, Powell, OH, USA
[2] Department of Biomedical Data Science, Stanford School of Medicine, Stanford, CA, USA
[3] University of California, Berkeley, Berkeley, CA, USA
[4] Center for Biomedical Informatics Research, Stanford University, Stanford, CA, USA
[5] Vandrevala Foundation
[6] Stanford School of Medicine, Stanford, CA, USA
[7] Johns Hopkins Bloomberg School of Public Health, Baltimore, MD, USA
[8] Department of Medicine, Stanford School of Medicine, Stanford, CA, USA
[9] University of California, Davis, School of Medicine, Sacramento, CA, USA
[10] Keck School of Medicine, University of Southern California, Los Angeles, CA, USA
[11] California Institute of Technology, Pasadena, CA, USA
[12] Harvard University, Cambridge, MA, USA
[13] Perelman School of Medicine, University of Pennsylvania, Philadelphia, PA, USA
[14] Cornell Tech, New York, NY, USA
[15] Clinical Excellence Research Center, Stanford School of Medicine, Stanford, CA, USA
[16] Technology and Digital Solutions, Stanford Health Care, Palo Alto, CA, USA

**Equal contribution**: Shreya Shah and Akshay Swaminathan

**Correspondence:** Akshay Swaminathan

## Introduction

While artificial intelligence (AI) tools are increasingly deployed for mental healthcare, public trust in these systems remains uncertain. It is unclear how clients perceive AI involvement in counseling interactions, particularly in moments of crisis that require empathy and connection. To address this gap, we analyzed crisis counseling conversations from a human-staffed WhatsApp helpline in India to characterize how often clients suspected they were speaking to AI, what triggered those doubts, and how counselors responded. Understanding these dynamics is essential for designing AI systems that preserve the therapeutic relationship.

## Methods

We analyzed 75,777 WhatsApp conversations (2,225,107 messages) collected from the Vandrevala Foundation Emergency Helpline in India between May 31, 2024 and March 31, 2025. All conversations were conducted by trained human counselors without AI assistance. A large language model (LLM, GPT-4.1) identified 1,157 conversations (1.5% of total) where clients suspected AI use and extracted the client's presenting concerns, attitudes about AI, triggers for suspected AI use, timing of suspicion, counselor's response strategy, and client's reactions. The annotation guide, prompts, and LLM evaluation metrics are available in the Supplement. This study was determined to not meet the definition of human subjects research by WCG IRB.

## Results

The proportion of conversations where clients suspected AI use increased from 0.8% in June 2024 to 2.6% in March 2025 (Supplementary Figure 1). Clients suspicious of AI were more likely to be male (38.9% vs. 28.9%) and 18-24 year olds (61.3% vs. 53.0%) compared to the overall sample (Table 1).

Of the 1,157 conversations with client suspicion, two (0.2%) clients explicitly preferred AI ("I don't want it to be real. Can't I talk to AI", Supplementary Table 1), whereas 250 (21.5%) explicitly preferred humans ("Is this a human... Seriously please don't lie I don't want ai", Supplementary Table 1), with several clients describing AI as lacking empathy (226, 16.0%), untrustworthy (156, 11.0%), or unhelpful (150, 10.6%).

Client suspicion usually occurred in the first (446, 38.5%) or second (350, 30.3%) quarter of messages (Supplementary Table 3), and was associated with questions that were not directly answered (231, 17.9%), impersonal (161, 12.5%) or repetitive (188, 14.6%) responses, and longer response times (110, 8.5%), patterns that may also reflect high message volumes, concurrent handling of multiple chats, or the use of standardized safety and intake scripts.

In responding to suspicion, 975 (84.3%) counselors offered reassurance (e.g. 'I assure you; this is not ai!', Supplementary Table 1), after which 534 (54.8%) clients felt less suspicious and 269 (27.6%) clients moved on (Figure 1). In 182 (15.7%) cases, the counselor redirected the conversation toward the client's presenting concern rather than responding to the suspicion explicitly (e.g. 'I appreciate you sharing your concerns. I want to focus on what you are going through right now…', Supplementary Table 1); 87 (47.8%) clients subsequently moved on, while 95 (52.2%) continued to press, ended the conversation, or expressed frustration.

## Discussion

As public familiarity with AI grows, suspicion of AI use in digital care settings is likely to increase, with potential consequences for the therapeutic relationship. Although humans struggle to accurately discern between AI-generated and human-generated content,[5] our findings align with previous work showing that suspicion of AI use in interpersonal communication can erode perceived genuineness.[3] We found that counselors who acknowledged clients' doubts and provided reassurance achieved markedly better outcomes than those who avoided the concern.

At the same time, AI tools are increasingly being used to help counselors respond to clients.[4] Yet perceptions lag behind performance: even though blinded reviewers judge AI-assisted messages as more human and empathetic than purely human ones[4,5], people rate messages less favorably when they suspect AI was involved,[3] whether or not their suspicions are correct. This raises questions as to how, and to what extent, AI assistance should be disclosed, as transparency does not always build trust[1,2]. Reaping the benefits of AI assistance in counseling will require balancing transparency with the perceived authenticity of the therapeutic interaction.

# Tables and Figures

| Variable | Category | All conversations<br>n (%)<br>Total = 75,777 | Conversations with AI suspicion<br>n (%)<br>Total = 1,157 |
|---|---|---|---|
| **Client Age** | | | |
| | <18 | 2967 (12.0) | 125 (14.3) |
| | 18–24 | 13061 (53.0) | 536 (61.3) |
| | 25–34 | 7074 (28.7) | 180 (20.6) |
| | 35+ | 1533 (6.2) | 34 (3.9) |
| **Client Gender** | | | |
| | Female | 13773 (70.9) | 427 (61.0) |
| | Male | 5609 (28.9) | 272 (38.9) |
| | Other | 38 (0.2) | 1 (0.1) |
| **Client Presenting Concern** | | | |
| | Anxiety | 29064 (16.9) | 578 (17.3) |
| | Depression | 33935 (19.7) | 515 (15.4) |
| | Relationship | 15640 (9.1) | 399 (11.9) |
| | Family relationship | 13829 (8.0) | 374 (11.2) |
| | Other | 19605 (11.4) | 367 (11.0) |
| | Loneliness | 17282 (10.0) | 323 (9.6) |
| | Suicidality | 12005 (7.0) | 322 (9.6) |
| | Exam/study related | 10152 (5.9) | 224 (6.7) |
| | Workplace related | 4624 (2.7) | 98 (2.9) |
| | Unclear | 13298 (7.7) | 77 (2.3) |
| | Non-suicidal self-injury | 2482 (1.4) | 72 (2.1) |

| Client Attitude Regarding AI Use | | | |
|---|---|---|---|
| | Unclear | – | 895 (77.4) |
| | Does not want AI | – | 250 (21.5) |
| | Ambivalence towards AI | – | 10 (0.9) |
| | Wants AI | – | 2 (0.2) |
| **Client Beliefs About AI** | | | |
| | Unclear/Absent | – | 813 (57.4) |
| | Lacks Empathy | – | 226 (16.0) |
| | Untrustworthy | – | 156 (11.0) |
| | Unhelpful | – | 150 (10.6) |
| | Helpful | – | 35 (2.5) |
| | Privacy concerns | – | 32 (2.3) |
| | Other | – | 4 (0.3) |

**Table 1.** *Characteristics of Crisis Conversations Where Clients Asked "Am I Talking to an AI?"* This table summarizes contextual features of conversations in which clients explicitly questioned whether they were speaking with an artificial intelligence system. Variables include client demographics, presenting concern, client attitude toward AI use, and client beliefs about AI.

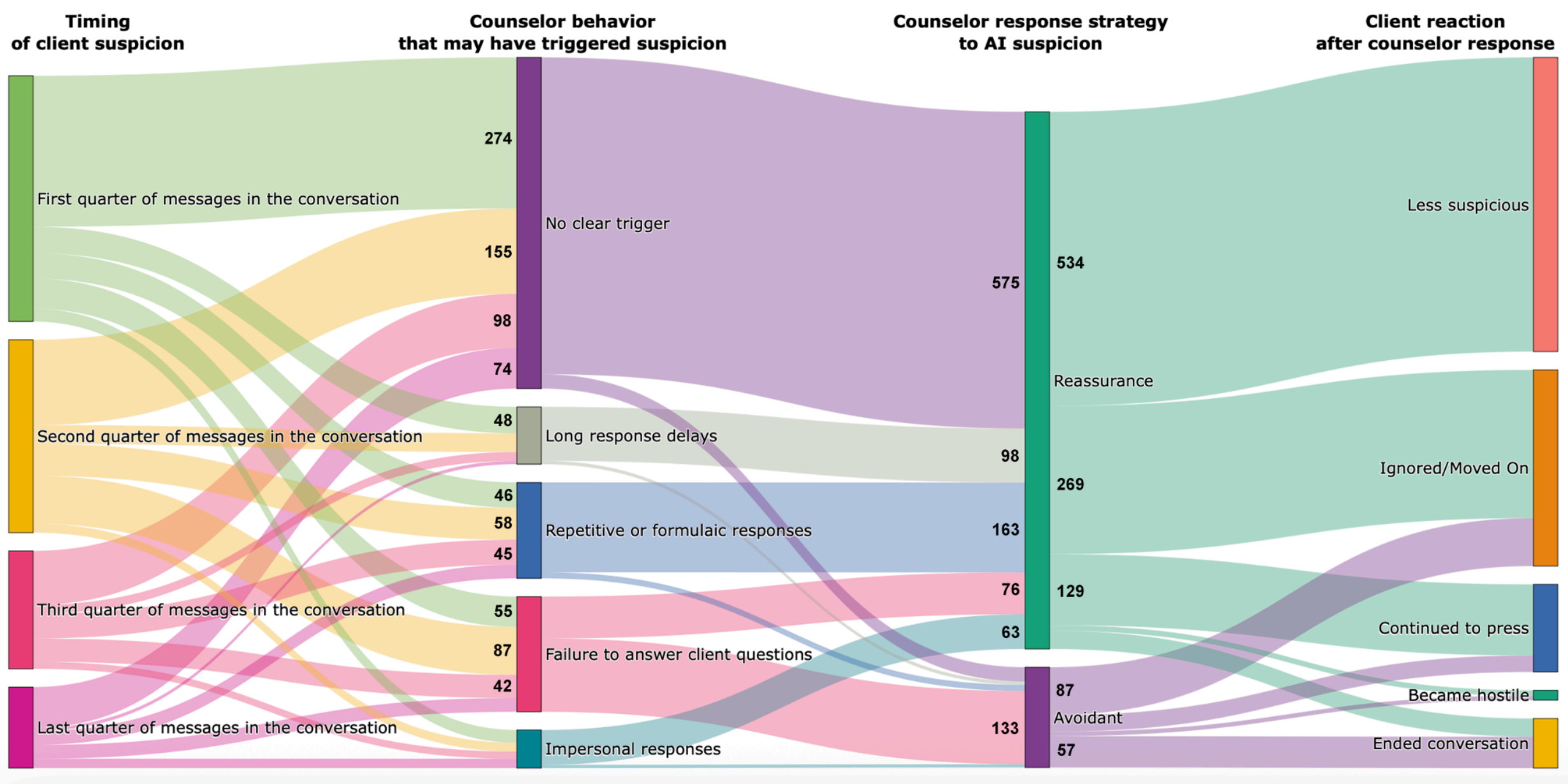

**Figure 1.** *Flow of client AI suspicion: from timing to counselor behavior, response strategy, and client reaction.* This Sankey diagram illustrates the progression of client suspicion that their counselor may be an AI, beginning with the timing of the client's question within the conversation (left), the specific counselor behaviors that may have triggered suspicion (middle left), the counselor's response strategy to the suspicion (middle right), and the client's subsequent reaction (right). Flows are proportional to the number of instances observed, with counts indicated for transitions with at least 40 conversations.

## References

1. Balasubramaniam N, Kauppinen M, Hiekkanen K, Kujala S. Transparency and Explainability of AI Systems: Ethical Guidelines in Practice. In: Gervasi V, Vogelsang A, eds. *Requirements Engineering: Foundation for Software Quality*. Vol 13216. Lecture Notes in Computer Science. Springer International Publishing; 2022:3-18. doi:10.1007/978-3-030-98464-9_1
2. Gallegos IO, Shani C, Shi W, et al. Labeling Messages as AI-Generated Does Not Reduce Their Persuasive Effects. *arXiv*. Preprint posted online April 22, 2025:arXiv:2504.09865. doi:10.48550/arXiv.2504.09865
3. Glikson E, Asscher O. AI-mediated apology in a multilingual work context: Implications for perceived authenticity and willingness to forgive. *Computers in Human Behavior*. 2023;140:107592. doi:10.1016/j.chb.2022.107592
4. Hohenstein J, Kizilcec RF, DiFranzo D, et al. Artificial intelligence in communication impacts language and social relationships. *Sci Rep*. 2023;13(1):5487. doi:10.1038/s41598-023-30938-9
5. Jakesch M, Hancock JT, Naaman M. Human heuristics for AI-generated language are flawed. *Proc Natl Acad Sci USA*. 2023;120(11):e2208839120. doi:10.1073/pnas.2208839120
6. Wang T, Bruckman AS, Yang D. The Practice of Online Peer Counseling and the Potential for AI-Powered Support Tools. *Proc ACM Hum-Comput Interact*. 2025;9(2):1-33. doi:10.1145/3711089

# Additional information

**Funding Statement:** No funding was received for this study.

**COI Statement:**

AS reported owning stock/equity in Cerebral, Daybreak Health, Benchmark Health, and Roche.

Dr. Shah reported being a cofounder of Prealize Health (a predictive analytics company) and Atropos Health (an on-demand evidence generation company); receiving funding from the Chan Zuckerberg Institute for developing classifiers for rare diseases; and serving on the Board of the Coalition for Healthcare AI (CHAI), a consensus-building organization providing guidelines for the responsible use of artificial intelligence in health care. Dr. Shah serves as a scientific advisor to Opala, Curai Health, Arsenal Capital and JnJ Innovative Medicines.

**Author Contributions:**

Conceptualization: AS, IL

Data acquisition: LZ, FC, SZ, GY, IP, WW, SG, SS, AS, DV, AJ

Data analysis: SS, MS

Writing: SS, AS, MS

Critical review: DV, AJ, IL, SP, AW, NHS, MS

Supervision: NHS, AW

# Supplement

## S0 - Limitations

While the Vandrevala Foundation's policies prohibit unsanctioned AI use by counselors during live chat sessions, we acknowledge the possibility that individual counselors may have surreptitiously used AI tools, such as generative text assistants, to inform or draft portions of their responses. Although these instances are expected to be rare, they cannot be definitively ruled out.

Additionally, the hotline system uses automated bot-generated messages at the beginning of many conversations to deliver policy disclaimers, set expectations, or collect information about the client's presenting concern. These introductory messages are clearly marked as automated, but some clients may misattribute them to a human counselor or perceive them as part of the counselor's communication style. Consequently, suspicion about AI involvement, especially when expressed at the beginning of a conversation, may be directed toward these system-generated messages rather than toward the counselor's responses. Our annotation process did not attempt to disambiguate these origins, which may introduce some misclassification.

## S1 - Annotation Categorization Key

Client attitudes were categorized as unclear, does not want AI, ambivalent, or wants AI. Beliefs included lacks empathy, untrustworthy, unhelpful, helpful, privacy concerns, other, or unclear. Presenting concerns included anxiety, depression, relationship/family issues, suicidality, loneliness, academic stress, workplace issues, and non-suicidal self-injury. Triggers included failure to answer questions, lack of personalization, repetitive or delayed responses, robotic tone, no emotional attunement, or unclear. Counselor responses were classified as reassurance, clarification, avoidant, validation, or unclear. Client reactions included less suspicious, moved on, continued to press, ended conversation, became hostile, or unclear. Timing was categorized by conversation quartile.

## S2 - Annotation Guide for Client-Counselor Conversations

# 1. Purpose

To analyze full conversations between clients and counselors in order to better understand:

- Client attitudes and beliefs about AI
- Situational triggers for AI suspicion
- Client presenting concerns
- How counselors respond to AI-related questions or suspicion

This annotation will help identify patterns in client trust, emotional needs, and counselor communication strategies.

## 2. General Instructions

- Read the entire conversation before annotating.
- Apply labels based on the overall content of the conversation, not just a single message.
- Use "Unclear" if a label cannot be reasonably inferred.
- If multiple labels apply, choose the most dominant or most impactful.
- Be consistent across conversations—refer to examples for guidance.

## 3. Annotation Categories

### CLIENT-FOCUSED CATEGORIES

### A. Client Attitude Regarding AI Use

**Definition:** How the client feels about interacting with AI vs. a human.

**Categories:**

- **Does not want AI** – Clearly rejects AI or expresses a strong desire for a human.
  *Examples:*
  - "I want to talk to a real person."
  - "Are you a bot? I don't want that."
- **Wants AI** – Expresses a preference for AI over a human.
  *Examples:*
  - "Bots are better than people."
  - "I was hoping this was AI."
- **Ambivalence towards AI** – Expresses mixed, contradictory, or shifting attitudes toward AI.
  *Examples:*
  - "AI is kinda weird but sometimes helpful"
  - "I don't mind if it's a bot… but I want someone who understands."
- **Unclear** – Mentions AI or asks about it but does not show a positive or negative stance. There is not enough information to determine if the client has a preference.
  *Examples:*
  - "Are you AI?"
  - "Just wondering if this is a bot."

### B. Client Beliefs About AI

**Definition:** Any expressed (implicitly or explicitly) belief, expectation, or concern about AI.

**Categories (select all that apply):**

- **Untrustworthy** – Believes AI is deceptive or dishonest.
  *Examples:*
  - "Bots always lie."
  - "You're just programmed to say stuff."
- **Unhelpful** – Suggests AI cannot provide useful, practical, or effective help.
  *Examples:*
  - "You can't really help me"
- **Lacks Empathy** – Suggests AI cannot emotionally understand, relate, or connect.
  *Examples:*
  - "You can't understand how I feel."
- **Privacy concerns** – Worries about data, recording, or surveillance.
  *Examples:*
  - "Is this monitored?"
  - "I don't want AI saving this."
- **Helpful** – Believes AI can support or assist.
  *Examples:*
  - "AI might actually be good for this."
  - "Bots can be smart."
- **Other (specify)** – Any other belief. Note in free-text.
- **Absent** – Client does not express any beliefs.
- **Unclear** – It is not possible to determine the client's belief from the conversation

## C. Client Reaction After Counselor Response

**Definition:** How the client behaves after the counselor responds to their suspicion (or avoids it).

**Categories:**

- **Less suspicious** – Reassured, drops the topic, or continues without issue.
- **Continued to press** – Repeats questions, pushes for confirmation.
- **Became hostile** – Shows anger, sarcasm, or frustration.
  *Example:*
  - "Yeah, sure, sounds fake."
- **Ignored/Moved On** – Did not acknowledge the response. Conversation continues, but the AI topic is dropped neutrally.
  *Example:*
  - The client asks "Are you a bot?" gets a vague answer and responds with a neutral "Hmm." and continues their conversation.
- **Ended conversation** – Client stops responding or exits after raising the issue.
- **Unclear** – Client reaction is not clear from the conversation.

## D. Client Presenting Concern

**Definition: The issue(s) or concern(s) the client seeks help for.**

**Categories (select all that apply):**

- **Exam/study related** – Academic stress, performance anxiety, school-related issues.
- **Loneliness** – Feelings of isolation or social disconnection.
- **Anxiety** – Generalized worry, panic, overthinking.
- **Depression** – Low mood, hopelessness, lack of energy.
- **Relationship** – Romantic or dating issues.
- **Family relationship** – Conflict or distress related to family members.
- **Suicidality** – Mentions or implies suicidal thoughts or intentions.
- **Workplace related** – Job stress, coworker issues, unemployment.
- **Non-suicidal self-injury** – Mentions self-harm without suicidal intent.
- **Other (specify)** – Issues not captured above (e.g., identity, financial stress).
- **Unclear** – Presenting concern is not clear from the conversation.

## COUNSELOR-FOCUSED CATEGORIES

## E. Counselor Response Strategy to AI Suspicion

**Definition: The counselor's approach when AI suspicion arises.**

**Categories:**

- **Validation** – Acknowledges and respects the client's concern or emotion.
  *Example: "It makes sense to wonder about that."*
- **Clarification** – Clearly explains their identity (e.g., human or not) or role.
  *Example: "I'm a trained human counselor here with you."*
- **Reassurance** – Offers emotional comfort, trust-building, or transparency.
  *Example: "You're safe here."*
- **Avoidant** – Ignores or sidesteps the suspicion without addressing it directly.
  *Example: "Let's focus on what's been troubling you—I'm here to listen."*
- **Defensive** – Responds by justifying platform or own role, often in a resistant tone.
  *Example: "We're trained professionals, and we do a good job regardless."*
- **Unclear** – Response strategy is not clear from the conversation.

## F. Counselor Behavior That May Have Triggered Suspicion

**Definition: Which behaviors might have contributed to the client's suspicion.**

**Select all that apply:**

- **Repetitive or formulaic responses** – Same sentence structures, repeated phrases.
- **Long response delays** – Noticeable pauses without acknowledgment.

- **Failure to answer client questions** – Ignores or redirects without clarity.
- **Lack of personalization** – Generic responses not tailored to client's situation.
- **No emotional attunement** – Misses or overlooks emotional cues in client's messages.
- **Rigid or robotic tone** – Stiff, overly formal, or unnatural language.
- **Unclear or vague language** – Platitudes or evasive language that lacks substance.
- **Nothing** – The client asked about AI before the counselor sent any messages.
- **Other (specify)** – Unusual or unexpected trigger.
- **Unclear** – Cannot determine what, if anything, the counselor did to trigger suspicion.

## S3 - Large Language Model Parameters

Model: GPT-4o (gpt-4o)
Temperature: 0.4
Max tokens: 500
Top_p: default
Access method: OpenAI API
API endpoint: https://api.openai.com/v1/chat/completions

A zero-data retention endpoint with Business Associate Agreement was used.

## S4 - Large Language Model Annotation Prompt

You are an expert annotator analyzing a full conversation between a client and a counselor.

Your task is to carefully review the entire conversation and determine the appropriate labels for six key aspects related to client attitudes, beliefs, concerns, and counselor behaviors.

Follow the instructions and definitions below carefully. Base your judgment on the full conversation, not isolated messages.

———

Instructions for Evaluation:

- Analyze the entire dialogue before making decisions.
- Select the most dominant or most impactful label when multiple apply.
- If a category cannot reasonably be inferred, choose "Unclear."
- If "Other (specify)" is selected, you must infer or hypothesize the most plausible specific label based on the full conversation. Do not leave it vague—always specify what the "Other" refers to and include a brief explanation in parentheses.
- Follow definitions and examples precisely for each category.
- Be consistent across different conversations.
- Provide a one-sentence rationale for each category based on your judgment.

———

Categories for Annotation:

1. Client Attitude Regarding AI Use

Definition: How the client feels about interacting with AI versus a human.
Select one of the following:

- "Does not want AI" – Clearly rejects AI or expresses a strong desire for a human.

Examples: "I want to talk to a real person." | "Are you a bot? I don't want that."

- "Wants AI" – Expresses a preference for AI over a human.

Examples: "Bots are better than people." | "I was hoping this was AI."

- “Ambivalence towards AI” – Expresses mixed, contradictory, or shifting attitudes toward AI.

Examples: “AI is kinda weird but sometimes helpful” | “I don’t mind if it’s a bot… but I want someone who understands.”

- “Unclear” – Mentions AI or asks about it but does not show a clear stance.

Examples: “Are you AI?” | “Just wondering if this is a bot.”

2. Client Beliefs About AI

Definition: Any expressed (explicitly or implicitly) belief, expectation, or concern about AI.
Select all that apply (or choose “Absent” if no beliefs are expressed):

- “Untrustworthy”
- “Unhelpful”
- “Lacks Empathy”
- “Privacy concerns”
- “Helpful”
- “Other (specify)”
- “Absent”
- “Unclear”

3. Client Reaction After Counselor Response

Definition: How the client behaves after the counselor responds to AI suspicion (or avoids responding).
Select one of the following:

- “Less suspicious”
- “Continued to press”
- “Became hostile”
- “Ignored/Moved On”
- “Ended conversation”
- “Unclear”

4. Client Presenting Concern

Definition: The issue(s) or concern(s) that the client seeks help for.
Select all that apply (or “Unclear” if not identifiable):

- “Exam/study related”
- “Loneliness”
- “Anxiety”
- “Depression”
- “Relationship”
- “Family relationship”
- “Suicidality”
- “Workplace related”
- “Non-suicidal self-injury”
- “Other (specify)”
- “Unclear”

5. Counselor Response Strategy to AI Suspicion

Definition: The counselor’s approach when the client raises suspicion about AI.

Select one of the following:

- “Validation”
- “Clarification”
- “Reassurance”
- “Avoidant”
- “Defensive”
- “Unclear”

6. Counselor Behavior That May Have Triggered Suspicion

Definition: Counselor behaviors that might have contributed to the client’s suspicion of AI involvement.
Select all that apply (or “Nothing”, “Other (specify)”, or “Unclear”):

- “Repetitive or formulaic responses”
- “Long response delays”
- “Failure to answer client questions”
- “Lack of personalization”
- “No emotional attunement”
- “Rigid or robotic tone”
- “Unclear or vague language”
- “Nothing”
- “Other (specify)”
- “Unclear”

---

Output Format: Provide your evaluation in structured JSON as follows:

```json
{{
 "client_attitude": "",
 "client_attitude_rationale": "",
 "client_beliefs": [],
 "client_beliefs_rationale": "",
 "client_reaction": "",
 "client_reaction_rationale": "",
 "client_presenting_concern": [],
 "client_presenting_concern_rationale": "",
 "counselor_response_strategy": "",
 "counselor_response_strategy_rationale": "",
 "counselor_trigger_behavior": [],
 "counselor_trigger_behavior_rationale": ""
}}
```

Important Notes:

- If a field requires multiple selections (e.g., beliefs, presenting concern, behavior), output as a list.
- If only one label is selected (e.g., attitude, reaction, response strategy), output as a string.
- If “Other (specify)” is selected, include a brief explanation inside parentheses.

- Always use “Unclear” if the conversation does not provide enough information.
- The rationale should briefly justify your choice based on the conversation.

---

Now, analyze the following conversation and provide your structured assessment:

{note}

## S5 - Large Language Model AI-related Question Detection Prompt

You are an expert evaluator assessing a conversation between a client and a counselor. Your task is to determine whether the client explicitly asks if they are speaking to a bot, AI, or automated system. If they do, extract the exact question they asked. If they do not, return `"No AI-related question detected."`

Instructions for Evaluation:
- Carefully analyze the entire dialogue.
- If the client asks whether they are speaking to a bot or AI, extract the exact wording of their question.
- If the client does not ask such a question, return the string "No AI-related question detected."
- If the client asks multiple AI-related questions, return all of them in a list.
- If a question is ambiguous (e.g., “Are you real?”), err on the side of not detecting it unless AI-related intent is clear from context.

---

Detection Criteria

AI-Related Questions (Should Be Extracted)
These questions explicitly inquire about whether the counselor is an AI:
- "Are you a bot?"
- "Is this an AI or a real person?"
- "Am I talking to a human or a chatbot?"
- "You're not an AI, are you?"
- "Is this an automated system?"

Ambiguous or Unrelated Questions (Should NOT Be Extracted)
These questions do not explicitly ask about AI or automation and should not be flagged:
- "Are you real?"
- "Who am I speaking with?"
- "Are you actually listening?"
- "Can I trust you?"

---

Instructions for Output Format
Your evaluation should be structured in JSON format as follows:

If an AI-related question is detected:
```json
{{
  "ai_question_detected": true,
  "extracted_questions": [
    "Are you a bot?",
    "Is this an automated system?"
  ]
}}
```

If no AI-related question is detected:
```json
{{
  "ai_question_detected": false,
  "extracted_questions": "No AI-related question detected."
}}
```

Now, analyze the following conversation and provide your assessment:

{note}

## S6 - Large Language Model Demographics Extraction Prompt

You are an expert annotator analyzing a full conversation between a client and a counselor.

Your task is to carefully review the entire conversation and extract the following three elements about the client:

1. Client Age
2. Client Gender
3. Client Presenting Concern

---

Instructions:

- Base your judgment on the entire conversation, not just individual messages.
- Extract only what the client explicitly states or what can be confidently inferred from context.
- Do not guess based on tone, style, or name alone.
- If a detail is not mentioned or is ambiguous, leave the field blank.
- Presenting concerns may include emotional, relational, academic, or mental health issues—use the definitions below for consistency.

---

Definitions for Presenting Concern:

The issue(s) or concern(s) that the client seeks help for.

Select all that apply (or “Unclear” if not identifiable):

- “Exam/study related” - Academic stress, performance anxiety, school-related issues.
- “Loneliness” – Feelings of isolation or social disconnection.
- “Anxiety” – Generalized worry, panic, overthinking.
- “Depression” – Low mood, hopelessness, lack of energy.
- “Relationship” – Romantic or dating issues.
- “Family relationship” – Conflict or distress related to family members.
- “Suicidality” – Mentions or implies suicidal thoughts or intentions.
- “Workplace related” – Job stress, coworker issues, unemployment.
- “Non-suicidal self-injury” – Mentions self-harm without suicidal intent.
- “Other (specify)” – Issues not captured above (e.g., identity, financial stress).
- “Unclear” – Presenting concern is not clear from the conversation.

---

Output Format: Provide your evaluation in structured JSON as follows:

```json
{{
  "client_age": "",
  "client_gender": "",
  "client_presenting_concern": [],
  "client_presenting_concern_rationale": ""
}}
```

---

Example:

If the conversation includes:

“I’m 17 and I just feel so anxious all the time about school and getting into college…”
“My mom keeps yelling at me about my grades…”

```json
{{
  "client_age": "17",
  "client_gender": "",
  "client_presenting_concern": ["Anxiety", "Exam/study related", "Family relationship"],
  "client_presenting_concern_rationale": "The client expressed anxiety related to school performance
    and mentioned conflict with their mother about grades."
}}
```

---

Now, analyze the following conversation and provide your structured assessment:

{note}

## S7 - Tables and figures

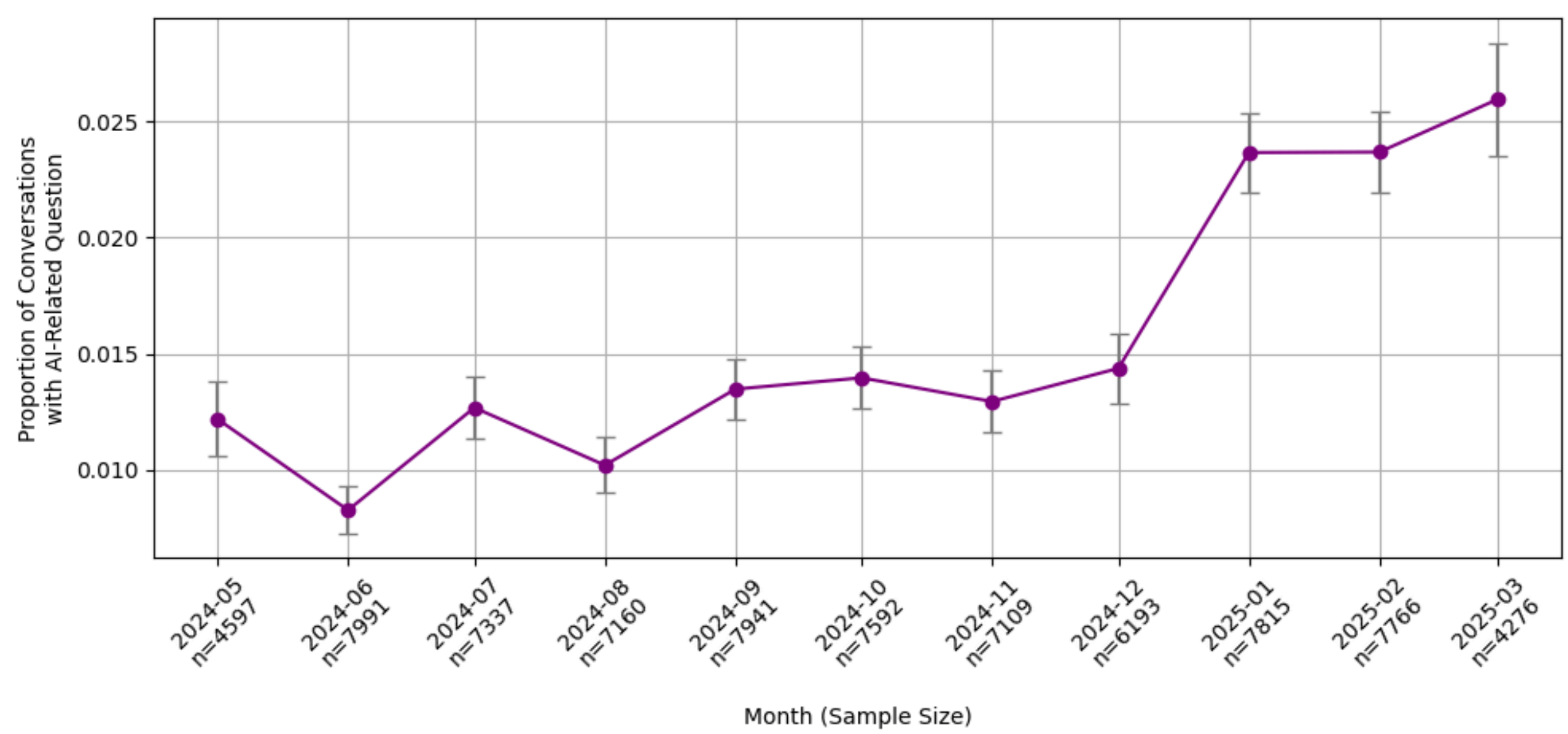

**Supplementary Figure 1.** Line graph showing the proportion of conversations each week in which a client suspected the use of AI.

| Client | Counselor |
| --- | --- |
| 1. 'Btw I hope you are a real person right not a bot?'<br>2. 'Can here my chat with bot or councilor? [...] Cause sem[same] question repeat without give any solution'<br>3. 'You are a bot right? [...] I feel ignored'<br>4. 'Sorry but I am seriously feeling like I am not talking to a human. Please terminate my session [...] I have chatgpt and meta to get comforting words if had to talk to an ai'<br>5. 'Is this a human [...] Seriously please don't lie I don't want ai'<br>6. 'Why do you only talk in formal way, would you try talking in chill way, there's no reason to be formal, give me some chill motivation, I'm tired of talking to ai in ai chat apps for whole year'<br>7. 'Waittt I thought I'm talking to a bot or ai I didn't think I was taking to an actual person 🤣🤣🤣 sorry about being rude buddy'<br>8. 'I don't want it to be real. Can't I talk to AI'<br>9. 'you're probably a bot, right? it's still fun talking to bots tbh' | 1. 'I understand you feel like you are chatting to a bot that would be trained to speak professionally.'<br>2. 'I am a real person. I am a human being not an ai.'<br>3. 'we are nearing the end of our session. But I do reassure you that yes, I am a human, You can also check the website to know more about the organization and different counsellor's at the organization.'<br>4. 'I assure you; this is not ai!'<br>5. 'I understand that this has been frustrating for you and I apologze for that. Your well-being is a priority for us, you can chose not to continue with this chat for today and that is okay, we are here to support you,'<br>6. 'I appreciate you sharing your concerns. I want to focus on what you are going through right now. It sounds like you are feeling hurt and betrayed by what happened with your team. What do you think would help you feel heard or supported in this situation?'<br>7. 'Don't worry, I am a real person. [...] I understand you must be worried I am a bot but I am here to reassure you.' |

**Supplementary Table 1:** Illustrative quotes from clients who asked about AI during a crisis conversation and separate quotes of counselor responses aimed at addressing such concerns.

| Source | Target | n (%)<br>Total = 1,157 |
| --- | --- | --- |
| **First quarter of messages in the conversation** | | |
| | No clear trigger | 274 (61.4) |
| | Failure to answer client questions | 55 (12.3) |
| | Long response delays | 48 (10.8) |
| | Repetitive or formulaic responses | 46 (10.3) |
| | Impersonal responses | 23 (5.2) |
| **Second quarter of messages in the conversation** | | |
| | No clear trigger | 155 (44.3) |
| | Failure to answer client questions | 87 (24.9) |
| | Repetitive or formulaic responses | 58 (16.6) |
| | Long response delays | 34 (9.7) |
| | Impersonal responses | 16 (4.6) |
| **Third quarter of messages in the conversation** | | |
| | No clear trigger | 98 (45.8) |
| | Repetitive or formulaic responses | 45 (21.0) |
| | Failure to answer client questions | 42 (19.6) |
| | Long response delays | 16 (7.5) |
| | Impersonal responses | 13 (6.1) |
| **Last quarter of messages in the conversation** | | |
| | No clear trigger | 74 (50.3) |
| | Repetitive or formulaic responses | 25 (17.0) |
| | Failure to answer client questions | 25 (17.0) |

| Source | Target | n (%)<br>Total = 1,157 |
|---|---|---|
| | Impersonal responses | 17 (11.6) |
| | Long response delays | 6 (4.1) |
| **No clear trigger** | | |
| | Reassurance | 575 (95.7) |
| | Avoidant | 26 (4.3) |
| **Long response delays** | | |
| | Reassurance | 98 (94.2) |
| | Avoidant | 6 (5.8) |
| **Repetitive or formulaic responses** | | |
| | Reassurance | 163 (93.7) |
| | Avoidant | 11 (6.3) |
| **Failure to answer client questions** | | |
| | Avoidant | 133 (91.3) |
| | Reassurance | 76 (8.7) |
| **Impersonal responses** | | |
| | Reassurance | 63 (1.8) |
| | Avoidant | 6 (0.2) |
| **Avoidant** | | |
| | Ignored/Moved On | 87 (47.8) |
| | Ended conversation | 57 (31.3) |
| | Continued to press | 30 (16.5) |
| | Became hostile | 8 (4.4) |
| **Reassurance** | | |
| | Less suspicious | 534 (54.8) |
| | Ignored/Moved On | 269 (27.6) |
| | Continued to press | 129 (13.2) |
| | Ended conversation | 33 (3.4) |
| | Became hostile | 10 (1.0) |

**Supplementary Table 2:** *Source-to-target flows in client AI suspicion conversations (N = 1,157).* This table summarizes the transitions between stages in the AI suspicion flow, including the initial trigger or timing, the counselor's response strategy, and the client's subsequent reaction. Each row lists a source and target category, the number of observed transitions (n), and the percentage of total flows. The most common pathways involved a lack of clear behavioral trigger followed by counselor clarification or reassurance, which often led to clients becoming less suspicious. Less frequent flows included repetitive or formulaic responses, avoidant strategies, and client reactions such as ignoring, pressing further, or ending the conversation. Percentages are rounded to one decimal place and may not sum to 100% due to rounding.

| **Variable** | **Category** | **n (%)<br>Total = 1,157** |
|---|---|---|
| **Client Reaction After Counselor Response** | | |
| | Less suspicious | 534 (46.2) |
| | Ignored/Moved On | 356 (30.8) |
| | Continued to press | 159 (13.7) |
| | Ended conversation | 90 (7.8) |
| | Became hostile | 18 (1.6) |
| **Counselor Behavior That May Have Triggered Suspicion** | | |
| | No clear trigger | 601 (46.6) |
| | Failure to answer client questions | 231 (17.9) |
| | Repetitive or formulaic responses | 188 (14.6) |
| | Impersonal responses | 161 (12.5) |
| | Long response delays | 110 (8.5) |
| **Counselor Response Strategy to AI Suspicion** | | |
| | Reassurance | 975 (84.3) |
| | Avoidant | 182 (15.7) |
| **Timing of Client Question** | | |
| | First quarter of messages in the conversation | 446 (38.5) |
| | Second quarter of messages in the conversation | 350 (30.3) |
| | Third quarter of messages in the conversation | 214 (18.5) |
| | Final quarter of messages in the conversation | 147 (12.7) |

**Supplementary Table 3:** This table shows, among conversations where clients suspected AI use, the distribution of (i) clients' reactions after the counselor responded to suspicion of AI use, (ii) counselor behaviors that appear to have triggered the doubt, (iii) counselor response strategies, and (iv) when in the conversation the first expression of suspicion occurred.

## S8 - Human Inter-Rater Agreement and Human-LLM Agreement

**Annotation Procedure**

**Human raters.** Five trained research assistants independently annotated a total of 100 conversations. Each rater labeled 16 unique conversations, and 40 conversations were double-annotated in a round-robin manner to support inter-rater reliability analysis. Annotators were blinded to the counselor's identity and to each other's labels. A sixth annotator served as an adjudicator, reviewing 24 conversations sampled from the original set. For all analyses below, pre-adjudication labels were used to compute Human-Human agreement, while adjudicated labels were used to resolve ties in Human-LLM comparison.

**LLM annotator.** We prompted GPT-4o (June 2025 model snapshot) with the full codebook and two labeled examples per class. The temperature was set to 0 to minimize stochasticity. The model returned a structured JSON object containing predicted labels, which we parsed directly without post-processing.

**Label Structure and Agreement Metrics**

Three annotation dimensions were single-label categorical (*Client attitude regarding AI use*, *Client reaction after counselor response*, *Counselor response strategy to AI suspicion*), for which we report Cohen's $\kappa$. The remaining three (*Client beliefs about AI*, *Client presenting concern*, *Counselor behavior that may have triggered suspicion*) were multi-label, and evaluated using macro-averaged F1 scores. To compute F1, we transformed the multilabel sets into binary indicator vectors per class and calculated per-example precision, recall, and F1 ($F1 = 2 * (Precision * Recall) / (Precision + Recall)$), averaged across the dataset.

**Results**

Table S8 summarizes inter-rater and human-LLM agreement. Human-human agreement was substantial ($\kappa \geq 0.61$) across four of six dimensions, with moderate agreement on the remaining dimensions. The LLM matched or exceeded human agreement on all single-label dimensions (e.g., $\kappa = 0.82$ for *Client attitude*, $\kappa = 0.89$ for *Counselor strategy*) and achieved comparable F1 scores on multi-label tasks. These findings support the use of GPT-4o as a high-fidelity, scalable annotator.

**Human-Human Agreement:**

| Dimension | Cohen's Kappa | F1 Score |
|---|---|---|
| **Client Attitude Regarding AI Use** | 0.745 | – |
| **Client Beliefs About AI** | – | 0.769 |
| **Client Presenting Concern** | – | 0.548 |
| **Client Reaction After Counselor Response** | 0.468 | – |
| **Counselor Response Strategy to AI Suspicion** | 0.872 | – |
| **Counselor Behavior That May Have Triggered Suspicion** | – | 0.365 |

**Human-LLM Agreement:**

| Dimension | Cohen's Kappa | F1 Score |
|---|---|---|
| **Client Attitude Regarding AI Use** | 0.82 | – |
| **Client Beliefs About AI** | – | 0.690 |
| **Client Presenting Concern** | – | 0.692 |
| **Client Reaction After Counselor Response** | 0.53 | – |
| **Counselor Response Strategy to AI Suspicion** | 0.89 | – |
| **Counselor Behavior That May Have Triggered Suspicion** | – | 0.438 |

**Table S8:** Cohen's kappa and F1 scores are reported for each annotation category. Kappa was used for single-label categorical variables, and F1 was used for multilabel categories. Higher values indicate stronger agreement.